\documentstyle[12pt]{article}
\input epsf.tex

\begin{document}

\baselineskip=15.5pt
\pagestyle{plain}
\setcounter{page}{1}

\begin{titlepage}

\begin{flushright}
PUPT-1738\\
hep-th/9710181
\end{flushright}
\vspace{20 mm}

\begin{center}
{\huge Born-Infeld String as a Boundary Conformal Field Theory}

\vspace{5mm}

\end{center}

\vspace{15 mm}

\begin{center}
L\'arus Thorlacius

\vspace{3mm}

Joseph Henry Laboratories\\
Princeton University\\
Princeton, New Jersey 08544
\end{center}

\vspace{1.5cm}

\begin{center}
{\large Abstract}
\end{center}

\noindent
Solutions of the Born-Infeld theory, representing strings extending 
from a Dirichlet $p$-brane, are also solutions of the higher derivative 
generalization of the Born-Infeld equations defining an exact open 
string vacuum configuration.  

\vspace{2cm}
\begin{flushleft}
October 1997
\end{flushleft}
\end{titlepage}
\newpage


\newcommand{\grad}{\nabla}
\newcommand{\tr}{\mathop{\rm tr}}
\newcommand{\half}{{1\over 2}}
\newcommand{\third}{{1\over 3}}
\newcommand{\be}{\begin{equation}}
\newcommand{\ee}{\end{equation}}
\newcommand{\bea}{\begin{eqnarray}}
\newcommand{\eea}{\end{eqnarray}}

\newcommand{\dint}[2]{\int\limits_{#1}^{#2}}
\newcommand{\D}{\displaystyle}
\newcommand{\PDT}[1]{\frac{\partial #1}{\partial t}}
\newcommand{\PD}{\partial}
\newcommand{\tw}{\tilde{w}}
\newcommand{\tg}{\tilde{g}}
\newcommand{\newcaption}[1]{\centerline{\parbox{6in}{\caption{#1}}}}

In recent work \cite{calmal,gibbons,howeetal}, Born-Infeld non-linear 
electrodynamics was shown to have solutions describing macroscopic
strings extending from a Dirichlet $p$-brane.  Fluctuations that
are transverse to both the $p$-brane and the string satisfy a wave 
equation that interpolates between the two-dimensional physics
of the string worldsheet and $p{+}1$-dimensional wave propagation
in the brane worldvolume \cite{calmal}.  This system was studied 
further in \cite{lpt}, where the Born-Infeld solutions were extended
to the case of strings emanating from multiple coincident $p$-branes
and the dynamics of transverse fluctuations compared to corresponding
supergravity calculations.  The gauge theory and supergravity 
calculations are in agreement even at parameter values for which 
derivatives of the gauge field strength are large in the region of 
interest.  This is surprising given that the Born-Infeld theory is only
the leading approximation, valid for slowly varying fields, to the open
string effective action, and is not expected to be reliable when field
gradients are large. 

In this paper we show that the Born-Infeld string solution in fact
satisfies the beta function equations that generalize Born-Infeld theory
to all orders in worldsheet perturbation theory.  In other words, it 
defines a boundary conformal field theory.  It is the only known exact
vacuum configuration of open superstring theory that has non-constant
gauge field strength.\footnote{There are known examples of bosonic
boundary conformal field theories with non-trivial scalar fields that 
correspond to periodic tachyon backgrounds \cite{ghm,poltho,bcft}.  
These models have either vanishing or constant gauge fields.}  

The worldvolume theory of a Dirichlet $p$-brane is a $p{+}1$-dimensional
$U(1)$ gauge theory coupled to scalar fields $\phi_\rho$ that describe
the transverse displacement of the brane worldvolume.  For slowly
varying fields it is given by the dimensional reduction of D=10
supersymmetric Born-Infeld electrodynamics: $F_{\alpha\beta}$ for
$\alpha,\beta=0,\ldots,p$ is the field strength of the abelian gauge 
field on the worldvolume; $F_{\alpha\rho}=\partial_\alpha\phi_\rho$ for
$\rho=p+1,\ldots,9$; and $F_{\rho\sigma}=0$ since the $\phi_\rho$ are 
only functions of worldvolume coordinates.  We will ignore
the gravitational backreaction due to the gauge field and
work in flat spacetime throughout.  The Born-Infeld equation of motion is 
\begin{equation}
\Bigl( {1\over \eta - F^2} \Bigr)_\lambda^{\ \nu}
\partial_\nu F_\mu^{\ \lambda} = 0 \ .
\label{bieq}
\end{equation}
The Born-Infeld string solution has only one of the scalar fields,
$\phi_9$, excited and it satisfies
\be
\partial_\alpha \phi_9 = \pm E_\alpha  \ ,
\ee
where $E_\alpha$ is the electric field due to a point charge (or more
generally a collection of point charges) in the $p$-brane worldvolume.
The choice of sign corresponds to a string extending in the positive
or negative $x^9$ direction.

It is convenient to combine the non-vanishing components of the 
gauge field and the scalar into a $p{+}2$-dimensional matrix, 
\begin{equation} F_m^{\ n} = \left[
\begin{array}{ccc}
0 & \chi {\vec{\partial}} \phi^9 & 0 \\
\chi {\vec{\partial}} \phi^9 & 0 & -{\vec{\partial}}\phi^9 \\
0 & +{\vec{\partial}} \phi^9 & 0 
\end{array}
\right] \  ,
\label{ansatz}
\end{equation}
where $\vec{\partial}=(\partial_1,\ldots,\partial_p)$ is the spatial 
gradient on the worldvolume, and $\chi=\pm 1$.  
For an ansatz of this form the Born-Infeld equation (\ref{bieq}) 
reduces to Laplace's equation,
\begin{equation}
{\vec{\partial}}^{\, 2}\phi_9 =0 \  ,
\label{scalarlap}
\end{equation}
whose general solution describes any number of strings attached to
the $p$-brane at various locations \cite{calmal,gibbons,howeetal}.  
These solutions are easily extended to the case of multiple coincident 
$p$-branes \cite{lpt}.  Then the worldvolume gauge theory is 
non-abelian but the string configurations only involve an abelian
subsector of the full theory.

All these string solutions have regions where the Born-Infeld
theory, which was used to derive them in the first place, breaks down.  
This is illustrated by the example of a single string extending
from $r=0$, where $r= \sqrt{\vec{x}^{\, 2}}$ is a radial worldvolume 
coordinate,
\begin{equation}
\phi_9(r) =  {b_p \over p-2} \, {1\over r^{p-2}} \  .
\label{coulomb}
\end{equation} 
Here $b_p$ is the unit of $U(1)$ charge in the worldvolume theory.
Near $r=0$ the field strength is large, but more significantly it has
large derivatives and is therefore not slowly varying.  In deriving the
Born-Infeld effective action one ignores contributions that involve
derivatives of $F_{\mu\nu}$ and this is only justified if
$\vert \grad^N F\vert << 1$ for all $N\geq 1$ (in string units where 
$\alpha'=1$).

Going beyond the Born-Infeld approximation involves the study of
open superstrings in a general abelian gauge field background.  In a
worldsheet sigma model approach this amounts to introducing a
supersymmetric Wilson line boundary interaction,
\be
S_A = {i\over 2\pi} \int_{\partial\Sigma} ds \left[
A_\mu(X){dX^\mu\over ds} 
-  {i\over 2}F_{\mu\nu}(X) \psi^\mu\psi^\nu \right] \ ,
\label{wilson}
\ee
and carrying out a systematic perturbative analysis \cite{andtse,clny}.
We find it convenient to use a worldsheet background field expansion,
$X^\mu =\bar{X}^\mu + \pi^\mu$.  The Wilson line gives rise to three
gauge invariant interaction vertices at $N$-th order in quantum fields,
where $N\geq 2$,
\bea
L_N^{(1)} &=& {1\over N!} \grad_{\mu_1}\ldots \grad_{\mu_{N-1}}
F_{\mu_N \nu}(\bar{X}) {d\bar{X}^\nu\over ds} 
\pi^{\mu_1}\ldots\pi^{\mu_N} \ , \nonumber \\
L_N^{(2)} &=& {N-1\over N!} \grad_{\mu_1}\ldots \grad_{\mu_{N-2}}
F_{\mu_{N-1}\mu_N}(\bar{X})  
\pi^{\mu_1}\ldots\pi^{\mu_{N-1}} {d\pi^{\mu_N}\over ds}\ ,  
\label{vertices} \\
L_N^{(3)} &=& {-i\over 2(N-2)!} \grad_{\mu_1}\ldots \grad_{\mu_{N-2}}
F_{\mu_{N-1} \mu_N}(\bar{X})  
\pi^{\mu_1}\ldots\pi^{\mu_{N-2}} 
\psi^{\mu_{N-1}} \psi^{\mu_N}
\ . \nonumber
\eea
The background field effective action is the sum of all 1PI vacuum 
diagrams for the fermions and $\pi$-fields \cite{dewabb}.  
In the absence of of closed string background fields, all interactions
take place at the worldsheet boundary and we only need to consider
the worldsheet propagators restricted to the boundary,
\be
\langle \pi^\mu(s) \pi^\nu (s') \rangle
= -2\eta^{\mu\nu} \log{\vert s-s'\vert}  \ ,\qquad
\langle \psi^\mu(s) \psi^\nu (s') \rangle
=-2i\eta^{\mu\nu} {1\over s-s' }  \ .
\ee
The perturbation expansion contains linearly and logarithmically
divergent diagrams that require regularization and renormalization.
The linear divergences will cancel due to worldsheet supersymmetry
but the logarithmic ones give rise to a non-trivial beta function 
$\beta_\mu^A(\bar{X})$ for
the Wilson line coupling, which must vanish for a consistent 
open-string vacuum configuration.  At one-loop order the vanishing
of the beta function is equivalent to the Born-Infeld equation 
(\ref{bieq}) \cite{acny,leigh}.  Beyond one loop the beta function
receives contributions involving higher derivatives of $F_{\mu\nu}$
in various combinations.  Explicit two-loop calculations in \cite{andtse}
showed that for open superstrings the Born-Infeld equation is in fact 
not corrected by terms with three derivatives, ({\it i.e.} 
$(\grad F)^3$, $\grad^2 F\, \grad F$, or $\grad^3 F$), but a
non-vanishing contribution involving five derivatives was 
identified.  There is every reason to expect further corrections at 
higher orders but fortunately their detailed structure will not be
required here.

Our proof that the ansatz (\ref{ansatz}) defines an exact open string
vacuum relies on simple algebraic properties of that particular 
field configuration, which ensure that all possible higher-order 
corrections to the Born-Infeld equation must in fact vanish.  
A similar line of reasoning has been used to show that
certain closed string backgrounds with a null symmetry, such as
plane-fronted waves, define exact conformal field theories \cite{akhs}.

Let us define for each $p$-vector $\vec{A}$ the $p{+}2$-dimensional
matrix 
\begin{equation} M(\vec{A}) = \left[
\begin{array}{ccc}
0 & \chi \vec{A} & 0 \\
\chi \vec{A} & 0 & -\vec{A} \\
0 & +\vec{A} & 0 
\end{array}
\right] \  ,
\label{mdef}
\end{equation}
with $\chi=\pm 1$.  Such matrices satisfy,
\bea
\tr \left[ M(\vec{A}) M(\vec{B}) \right] &=& 0 \ ,
\label{tracetwo} \\
M(\vec{A}) M(\vec{B}) M(\vec{C}) &=& 0 \ ,
\label{nilpot}
\eea
for any vectors $\vec{A}$, $\vec{B}$, and $\vec{C}$.

The field strength tensor (\ref{ansatz}) and all its derivatives
$\grad_{\mu_1}\ldots \grad_{\mu_N} F_m^{\ n}$ are matrices of this
type and this allows us to make statements about general diagrams
in the perturbation expansion.  Any graph that contributes to the
gauge field beta function will have precisely one vertex of type-1
in (\ref{vertices}).  All other vertices in the 
diagram are either of type-2 or type-3 and they combine
in such a way that all fermions and $\pi$-fields are contracted, for
otherwise this would not be a vacuum diagram for the quantum fields.
An example of a diagram of this type is shown in Figure~1.

\vskip .5cm
\vbox{
{\centerline{\epsfxsize=2.4in \epsfbox{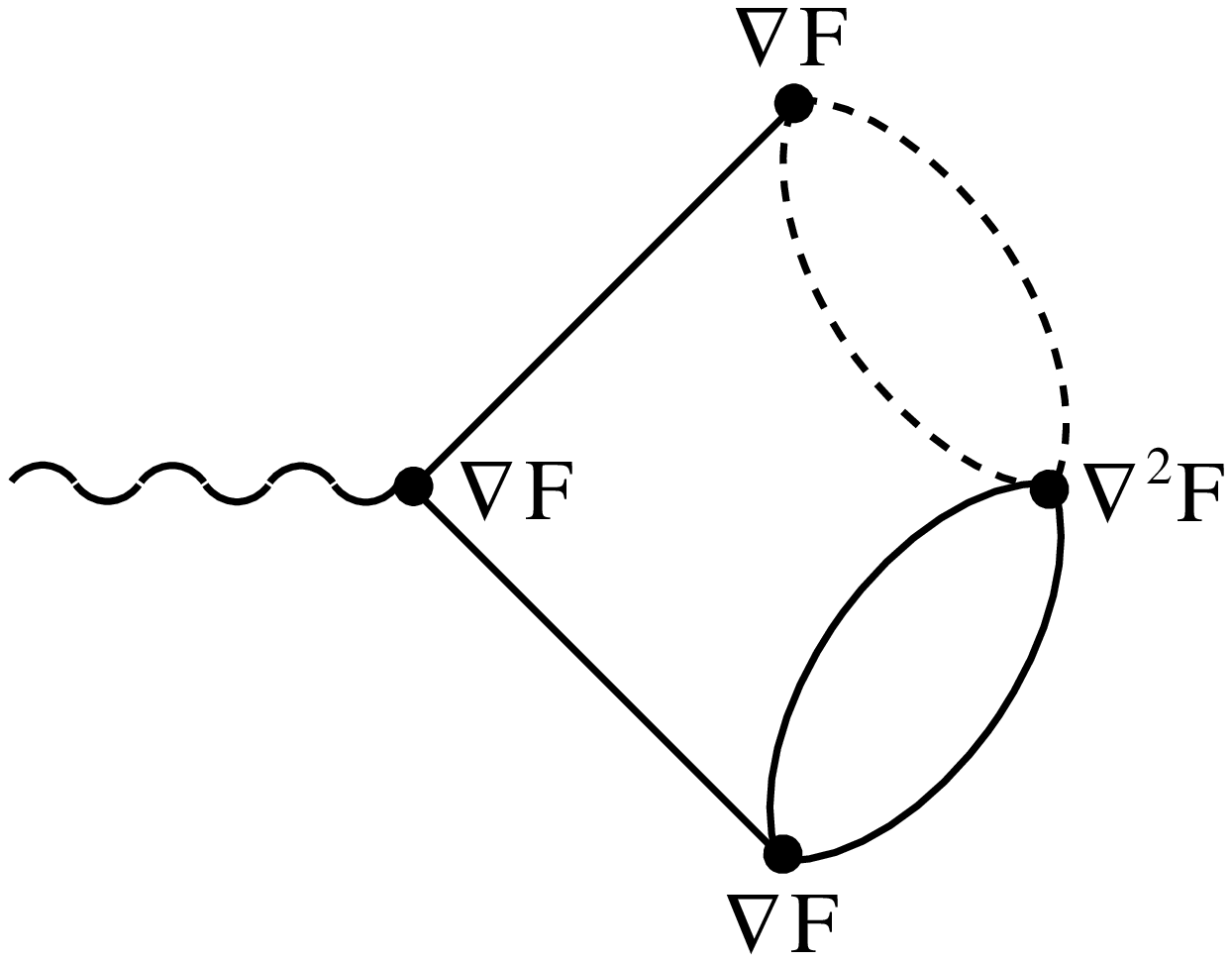}}}
{\centerline{ FIGURE 1:
A diagram that contributes to $\beta_\mu^A(\bar{X})$.  Dashed}}
{\centerline{lines are fermion propagators, solid lines 
$\pi^\mu$ propagators, }}
{\centerline{and the wavy line denotes the background field
${d\bar{X}^\nu \over ds}$.}}
}
\vskip .5cm

Now, let us analyze an arbitrary diagram that contributes to 
$\beta_\mu^A(\bar{X})$.  We first observe that any diagram with 
vertices of type-3 in (\ref{vertices}), {\it i.e.} with one or more 
closed fermion loops, 
vanishes for our ansatz.  This is because the fermion propagators 
contract the indices of the field strengths at the type-3 vertices so 
that we have a trace.  Each field strength will in general have some 
number of derivatives acting on it but that is still a matrix of the 
form (\ref{mdef}).  It follows from (\ref{tracetwo}) and (\ref{nilpot}) 
that the trace of a product of any number of such matrices vanishes.

We therefore only need to consider purely bosonic diagrams with 
one vertex of type-1 in (\ref{vertices}) and any number of type-2
vertices.  We proceed in a few steps, starting at the
type-1 vertex, which contributes the factor
\be 
\grad_{\mu_1}\ldots \grad_{\mu_{N-1}}
F_{\mu_N \nu}(\bar{X}) {d\bar{X}^\nu\over ds}  \ ,
\ee
for some value of $N\geq 2$.  Note that the ${d\bar{X}^\nu\over ds}$ is
contracted against an index on the field strength and not one of the 
derivatives.  We now follow the internal leg of the diagram that contracts
onto the other index $\mu_N$, of the field strength.  This leads us to
one of two possibilities: (i) The leg could contract back onto
the type-1 vertex; or (ii) contract onto one of the type-2 vertices.
Let us consider each case in turn.

(i) In this case we have a derivative contracted onto the field strength,
$\grad{\ldots}\grad_{\mu_N}F^{\mu_N \nu}$, and this vanishes for our
ansatz (\ref{ansatz}) when $\phi_9$ satisfies the Laplace equation
(\ref{scalarlap}).

(ii) Here our next step depends on whether the leg we are following
contracts:  (a) onto an index on one of the derivatives acting on the 
field strength $F$ at the second vertex; or (b) onto an index of $F$ 
itself.

In case (ii-a) we can use the Bianchi identity
\be
\grad_\mu F_{\nu\lambda}+\grad_\nu F_{\lambda\mu}
+\grad_\lambda F_{\mu\nu} = 0 \ ,
\ee
to `move' the derivative index onto the field strength, thereby reducing
this case to (ii-b) which we discuss next.

In case (ii-b) we consider the remaining index of the field strength at 
the second vertex.  If we follow the leg of the diagram that contracts
onto this index we are led to three distinct possibilities:
(1) The leg contracts back onto the same vertex; (2) it contracts onto
a third vertex; or (3) it contracts back onto the original type-1 vertex.  
Again we discuss each case in turn.

(ii-b-1)  The diagram vanishes for the same reason as in case (i).

(ii-b-2)  Here there are two possibilities.  The second internal leg either
contracts onto an index of the field strength at the third vertex or onto 
an index of a derivative acting on $F$.  In the latter case we can use the
Bianchi identity to move the index from the derivative to $F$ itself
and proceed from there.  When the leg contracts onto an index of $F$ 
we have a sequence of three matrices of the form (\ref{mdef})
contracted together and this vanishes by the nilpotence property
(\ref{nilpot}).  

(ii-b-3) Finally we have to consider the case where the second leg 
leads back to the first type-1 vertex.  We know that it must contract
onto an index of a derivative there since both indices of the field strength
are already accounted for.  By using the Bianchi identity at the type-1
vertex, combined with the antisymmetry of $F$ at the second vertex,
we can move the index that contracts with the background field
${d\bar{X}^\nu\over ds}$ onto the derivative of the field strength at
the type-1 vertex.  The diagram then contains a factor of the form,
\be
\ldots(\grad_{\mu_1}\ldots\grad_{\nu_1}F_{\nu_2\nu_3}
{d\bar{X}^{\nu_1}\over ds}) \> (\grad_{\lambda_1}\ldots
F_{\nu_2\nu_3})\ldots \ ,
\ee
which means that it vanishes by the trace property (\ref{tracetwo}).

We have shown that all open string tree diagrams that could 
contribute to the Wilson line beta function $\beta_\mu^A(\bar{X})$ 
vanish for field configurations that describe Born-Infeld strings.  
The argument relies on the special algebraic form of the Born-Infeld 
string solution but is independent of the renormalization scheme 
used to define the beta function.  
Since the diagrams involving fermions vanish separately from
the diagrams involving only bosons, it is clear
that the Born-Infeld string is also an exact solution of the bosonic
open string theory.  Our analysis is limited to 
open string theory at the classical level.  It does not include
closed string backgrounds and no open string loop diagrams.

We have chosen to look at the spacetime equations of motion of the
gauge field rather than its supersymmetry properties.  Our success
in showing that they are satisfied to all orders in worldsheet 
perturbation theory is nevertheless no doubt due to the fact that 
these are supersymmetric configurations.  In fact the ansatz, 
$\partial_\alpha \phi_9=\pm E_\alpha$, was originally arrived at
from BPS considerations in the linearized Maxwell approximation
to the Born-Infeld theory \cite{calmal,gibbons}.  An alternative 
approach to the one presented here would be to investigate higher 
order corrections to supersymmetry variations, and show that they 
all vanish for Born-Infeld strings.

\section*{Acknowledgements}

The author wishes to thank C. Callan, S. Lee, A. Peet, and W. Taylor 
for useful discussions.  This work was supported in part by a US 
Department of Energy Outstanding Junior Investigator Award, 
DE-FG02-91ER40671.

\end{document}